\documentclass[12pt]{article}
\usepackage{graphicx}
\newcommand{\ba}{\begin{array}}
\newcommand{\ea}{\end{array}}
\newcommand{\nn}{\nonumber\\}
\newcommand{\R}{{\bf R}}
\newcommand{\C}{{\bf C}}
\newcommand{\cH}{{\cal H}}
\newcommand{\del}{\partial}
\newcommand{\bra}{\langle}
\newcommand{\ket}{\rangle}
\newcommand{\Bra}{\Big{\langle}}
\newcommand{\Ket}{\Big{\rangle}}
\newcommand{\vvert}{\Big{\vert}}
\newcommand{\rar}{\rightarrow}
\newcommand{\lar}{\leftarrow}

\newcommand{\dar}{\downarrow}

\newcommand{\unl}{\underline}
\newcommand{\fr}{\frac}
\newcommand{\half}{\frac{1}{2}}

\newcommand{\scr}{\scriptsize}
\newcommand{\dis}{\displaystyle}

\newcommand{\zb}{\bar{z}}
\newcommand{\ah}{{\hat{a}}}
\newcommand{\fh}{{\hat{f}}}
\newcommand{\nh}{{\hat{n}}}
\newcommand{\xh}{{\hat{x}}}
\newcommand{\zh}{{\hat{z}}}
\newcommand{\zbh}{{\hat{\bar{z}}}}

\newcommand{\delh}{{\hat{\partial}}}
\newcommand{\unb}{\underbrace}
\makeatletter
\@addtoreset{equation}{section}

\makeatother

\begin{document}

\begin{titlepage}
\null
\begin{flushright}
UT-02-56
\\
hep-th/0301213
\\
January, 2003
\end{flushright}

\vskip 1.5cm
\begin{center}

  {\LARGE \bf Noncommutative Burgers Equation}

\lineskip .75em
\vskip 2cm
\normalsize

 {\large Masashi Hamanaka\footnote{On leave of absence
from university of Tokyo, Hongo.
New e-mail: hamanaka@hep1.c.u-tokyo.ac.jp}}
and
 {\large Kouichi Toda\footnote{e-mail:
kouichi@yukawa.kyoto-u.ac.jp}}

\vskip 2cm

  ${}^1${\it Institute of Physics, University of Tokyo, Komaba,\\
              Meguro-ku, Tokyo 153-8902, Japan}

\vskip 0.5cm

  ${}^2${\it Department of Mathematical Physics,
Toyama Prefectural University,\\
Toyama, 939-0398, Japan}

\vskip 1.5cm

{\bf Abstract}

\end{center}

We present a noncommutative version of the Burgers equation 
which possesses the Lax representation
and discuss the integrability in detail.
We find a noncommutative version of the Cole-Hopf transformation
and succeed in the linearization of it.
The linearized equation is the (noncommutative)
diffusion equation and exactly solved.
We also discuss the properties of some exact solutions.
The result shows that 
the noncommutative Burgers equation is completely integrable
even though it contains infinite number of time derivatives.
Furthermore, we derive the noncommutative Burgers equation from
the noncommutative (anti-)self-dual Yang-Mills equation
by reduction, which is an evidence for
the noncommutative Ward conjecture.
Finally, we present a noncommutative version of the Burgers hierarchy
by both the Lax-pair generating technique and the Sato's approach.

\end{titlepage}
\clearpage
\baselineskip 5.9mm

\section{Introduction}

The extension of ordinary integrable systems
to noncommutative (NC) spaces
has been studied intensively for the last several years
\cite{DiMH}-\cite{CFZ}.
In the recent developments of NC field theories,
various new physical aspects of gauge theories 
were revealed \cite{NC}
and several long-standing problems in real physics 
were solved.

NC field theories can be described 
as deformed theories from commutative ones.
In terms of gauge theories,
the deformation is essentially unique because
it corresponds to the presence
of the background magnetic fields \cite{NC}.
Among them, NC (anti-)self-dual Yang-Mills
equations are integrable and important \cite{YM}.
The first breakthrough was the great work \cite{NeSc}
of Atiyah-Drinfeld-Hitchin-Manin
construction \cite{ADHM} of NC $U(1)$ instantons.

On the other hand,
in the lower-dimensional theories,
there are many typical integrable equations
such as the Korteweg-de Vries (KdV) equation \cite{KdV}.
These equations contain no gauge field and
the NC extension of them perhaps might have  
no physical picture.
Furthermore, the NC extension of $(1+1)$-dimensional equations
introduces infinite number of time derivatives
and it becomes very hard to define the integrability.

Nevertheless, NC versions of them have been proposed
in various contexts.
They actually possess some integrable properties, 
such as the existence of infinite number 
of conserved quantities \cite{DiMH, GrPe}.
Furthermore, a few of them can be derived from 
NC (anti-)self-dual Yang-Mills equations
by suitable reductions \cite{Legare}.
This fact may give some physical meanings
to the lower-dimensional NC field equations.
Now it is time to investigate various aspects of them more in detail
in order to confirm whether the NC field
equations presented are really integrable or not.


\vspace{3mm}

For this purpose,
the Burgers equation \cite{Burgers} would be the best example.
On the commutative space-time,
it can be derived from the Navier-Stokes equation
and describes real phenomena, such as the turbulence 
and shock waves.
In this point, the Burgers equation draws much attention
amongst many integrable equations.
Furthermore, it can be linearized by
the Cole-Hopf transformation \cite{CoHo}.
The linearized equation is the diffusion equation and
can be solved by Fourier transformation for given boundary conditions.
This shows that the Burgers equation is completely integrable.
The Burgers equation actually sat in the central position
at the early stage of integrable systems
and have given much influence on the subsequent studies.
For example, the Hirota's bilinear transformation \cite{Hirota},
which is a simple generalization of the Cole-Hopf transformation,
plays a crucial role in the construction of the exact multi-soliton
solution of various soliton equations.
Therefore if the NC Burgers equation is
linearizable and integrable even on the NC space-time,
it would be the first example of completely integrable NC equations and 
has much significance for further studies on the topics.

\vspace{3mm}

In this article, we present NC versions
of the Burgers equation and the Burgers hierarchy
which possess the Lax representations.
We prove that the NC Burgers equation
is linearizable by a NC version of the Cole-Hopf transformation.
This shows that the NC Burgers equation 
is really integrable even though the
NC Burgers equation contains infinite number
of time derivatives in the nonlinear term.
The linearized equation is the (NC)
diffusion equation of first order with respect to time
and the initial value problem is well-defined.
The NC Lax representation is derived from
the compatibility condition of NC versions of linear systems. 
Hence the integrability of the NC Burgers equation 
with the Lax representation could relate to some symmetry and 
the existence of the NC Burgers hierarchy
might suggest a hidden infinite-dimensional symmetry
which is considered as a deformed symmetry from commutative one.
We also obtain the exact solutions 
which actually reflect the effects of the NC deformation.
Finally we derive the NC Burgers equation
from both NC (anti-)self-dual Yang-Mills equation
and the framework of NC extension of Sato theory.
This would be the first step to the confirmation of 
NC Ward conjecture and the completion of NC Sato theory.
We also discuss general properties of integrability in
NC field theories mainly in section 3 and 6.

\vskip5mm\noindent
{\bf Note added:}
After submitting
the present article and our paper \cite{HaTo}, 
we were aware of the paper
\cite{MaPa} by L. Martina and O. K. Pashaev
on arXiv e-print server,
which contains some overlaps with ours.

\section{Review of Noncommutative Field Theories}

NC spaces are defined
by the noncommutativity of the coordinates:
\begin{eqnarray}
\label{nc_coord}
[x^i,x^j]=i\theta^{ij},
\end{eqnarray}
where $\theta^{ij}$ are real constants and  
called the {\it NC parameters}.
This relation looks like the canonical commutation
relation in quantum mechanics
and leads to ``space-space uncertainty relation.''
Hence the singularity which exists on commutative spaces
could resolve on NC spaces.
This is one of the prominent features of NC
field theories and yields various new physical objects.

NC field theories 
have the following two equivalent descriptions:
\begin{itemize}
\item the star-product formalism
\item the operator formalism
\end{itemize}
These are connected one-to-one by the Weyl transformation,
which is briefly summarized in Appendix A.
In the present article, we mainly use the star-product formalism.

\vspace{2mm}
\noindent
\unl{\it The star-product formalism}
\vspace{2mm}

The star-product is defined for ordinary fields on commutative spaces.
For Euclidean spaces, it is explicitly given by 
\begin{eqnarray}
f\star g(x)&:=&
\exp{\left(\frac{i}{2}\theta^{ij}\partial^{(x^\prime)}_i
\partial^{(x^{\prime\prime})}_j\right)}
f(x^\prime)g(x^{\prime\prime})\Big{\vert}_{x^{\prime}
=x^{\prime\prime}=x}\nonumber\\
&=&f(x)g(x)+\frac{i}{2}\theta^{ij}\partial_if(x)\partial_jg(x)
+{\cal O}(\theta^2),
\label{star}
\end{eqnarray}
where $\del_i^{(x^\prime)}:=\del/\del x^{\prime i}$
and so on.
This explicit representation is known 
as the {\it Moyal product} \cite{Moyal}.

The star-product has associativity: $f\star(g\star h)=(f\star g)\star h$,
and returns back to the ordinary product 
in the commutative limit: $\theta^{ij}\rar 0$.
The modification of the product  makes the ordinary 
spatial coordinate ``noncommutative,'' 
that is, $[x^i,x^j]_\star:=x^i\star x^j-x^j\star x^i=i\theta^{ij}$.

NC field theories are given by the exchange of
ordinary products in the commutative field theories for
the star-products
and realized as deformed theories from the commutative ones.
In this context, they are often called the {\it NC-deformed theories}.

We note that the fields themselves take c-number values 
as usual and the differentiation and the integration for them 
are well-defined as usual.
A nontrivial point is that
NC field equations contain infinite number of
derivatives in general. Hence the integrability of the equations
are not so trivial as commutative cases.

\vspace{2mm}
\noindent
\unl{\it The operator formalism}
\vspace{2mm}

In order to make some comments on 
the integrability of NC field equations later,
let us introduce another formalism, the {\it operator formalism},
which is equivalent to the star-product formalism.

This time, we start with the noncommutativity of the
spatial coordinates
(\ref{nc_coord}) and define NC gauge theory
considering 
the coordinates as operators. 
{}From now on, we write the hats on the fields 
in order to emphasize that they are operators.
For simplicity, we treat NC plane
with the coordinates $\xh^1,\xh^2$ 
which satisfy $[\xh^1,\xh^2]=i\theta,~\theta>0$.

Defining new variables $\ah,\ah^\dagger$ as
\begin{eqnarray}
\ah:=\fr{1}{\sqrt{2\theta}}\zh,~\ah^\dagger:=\fr{1}{\sqrt{2\theta}}\zbh,
\end{eqnarray}
where $\zh=\xh^1+i\xh^2,~\zbh=\xh^1-i\xh^2$,
we get the Heisenberg's commutation relation:
\begin{eqnarray}
\label{heisenberg}
{[\ah,\ah^\dagger]}=1.
\end{eqnarray}
Hence the spatial coordinates can be considered 
as the operators acting on
Fock space $\cH$ which is spanned 
by the occupation number basis $\vert
n\ket:=\left\{(\ah^\dagger)^n/\sqrt{n!}\right\}\vert 0\ket,
~\ah\vert 0\ket=0$:
\begin{eqnarray}
\label{fock}
\cH=\oplus_{n=0}^{\infty}\C\vert n\ket.
\end{eqnarray}
Fields on the space depend on the spatial coordinates 
and are also the operators
acting on the Fock space $\cH$. 
They are represented by the occupation number basis as
\begin{eqnarray}
\fh=\sum_{m,n=0}^{\infty}f_{mn}\vert m\ket\bra n\vert.
\end{eqnarray}
If the fields have rotational symmetry on the plane,
that is, the fields commute with 
the number operator $\nh:=\ah^\dagger \ah\sim (\xh^1)^2+(\xh^2)^2$,
they become diagonal:
\begin{eqnarray}
\fh=\sum_{n=0}^{\infty}f_{n}\vert n\ket\bra n\vert.
\end{eqnarray} 

The derivation is defined as follows:
\begin{eqnarray}
\del_i\fh :=[\delh_i,\fh] := [-i(\theta^{-1})_{ij}\xh^j,\fh],
\end{eqnarray} 
which satisfies the Leibniz rule and the desired relation:
\begin{eqnarray}
\del_i\xh^j=[-i(\theta^{-1})_{ik}\xh^k,\xh^j]=\delta_i^{~j}.
\end{eqnarray}
The operator $\delh_i$ is called the {\it derivative operator}.
Hence we can define ``differential equations''
which are realized as recursion relations 
for the matrix element $f_{mn}$.

The integration can also be defined as the trace of the
Fock space $\cH$:
\begin{eqnarray}
\int dx^1dx^2~ \fh(\xh^1,\xh^2)&:=& 2\pi \theta
{\mbox{Tr}}_\cH \fh.
\end{eqnarray}
Hence the solutions for 
``differential equations'' are also well-defined.

\section{Comments on Integrability of Noncommutative Field Equations}

Before NC extension of the Burgers equation,
let us discuss what is the definition of integrability
of NC field equations.

Even on commutative spaces,
it is hard to define what 
is integrability of field equations.
(See e.g. \cite{Zakharov}.)
There are various definitions for it according to situations.
Typical definitions are as follows.
Field equations are called integrable
if they possess, for example, 
the linearizability,
the Lax representation, 
the existence of infinite number of conserved quantities,
the bi-Hamiltonian structure,
the exact multi-soliton solutions,
the Painlev\'e properties,
the presence of algebraic geometry,
and so on. (See e.g. \cite{FaTa, AbCl})
Among them, the linearizability is the best definition
because the linearized equation can be solved by
Fourier transformation for arbitrary given initial conditions.

On NC spaces, it becomes harder to define what 
is integrability of field equations.
In this case, there are two situations 
which should be discussed separately:
\begin{itemize}
 \item space-space noncommutativity
 \item space-time noncommutativity
\end{itemize} 

In the former, the situation is just the same as
the ordinary commutative case because NC field theories
can be considered as just deformed theories. 
The fields are c-number valued functions
(or infinite-side matrices in the operator formalism)
and the derivation and the integration is well-defined as usual.
The notions of 
time evolutions,
Hamiltonian structures,
action-angle variables
and inverse scattering methods
are also well-defined.
Hence the above definitions for commutative field equations
are also reasonable for those NC equations 
with space-space noncommutativities.

In the latter, however, the situation changes drastically.
The obstruction arises in the notion of time evolution
in nonlinear equations. 
For simplicity, let us consider the ($1+1$)-dimensional NC space-time
whose coordinates and noncommutativity
are $(x,t)$ and $[t,x]=i\theta$, respectively. 
The noncommutativity
introduces infinite number of time derivatives 
in nonlinear terms as
\begin{eqnarray}
\label{star_11}
f\star g(t,x)=
e^{\frac{i}{2}\theta
\left(
\partial_{t^\prime}
\partial_{x^{\prime\prime}}
-\partial_{x^\prime}
\partial_{t^{\prime\prime}}\right)}
f(t^\prime,x^\prime)g(t^{\prime\prime},x^{\prime\prime})
\Big{\vert}_{\scr
\ba{c}
t^{\prime}
=t^{\prime\prime}=t\\
x^{\prime}
=x^{\prime\prime}=x,
\ea}
\end{eqnarray}
where $\del_t=\del / \del t$ and so on.
Hence the notion of time evolution becomes vague and 
the infinite number of derivatives of time
might lead to acausal structure into the theories.
The initial value problem also seems to be hard to define.
Therefore it becomes seriously hard to discuss the integrability.
In this case, only one possible definition of integrability
is the linearizability because linearized equations
contain neither nonlinear term nor the star-product. 

The Burgers equation is defined on $(1+1)$-dimensional
space-time and the NC extension belongs to the latter case.
Hence it is worth studying whether it is linearizable
and how the initial value problem is solved.

\section{Noncommutative Burgers Equation}

In order to do it,
we first present some NC version of the Burgers equation
on $(1+1)$-dimensional noncommutative space-time 
with the noncommutativity: $[t,x]=i\theta$.
In this section,
we construct a NC version of the Burgers equation
which possesses the Lax representation.

A given NC differential equation is said to have
the Lax representation
if there exists a suitable pair of operators $(L,T)$
so that the following equation (the {\it Lax equation})
\begin{eqnarray}
\label{lax}
[L,T+\del_t]_\star=0
\end{eqnarray}
is equivalent to the given NC differential equation.
Here the star-product does not affect the derivative operator,
for example, $\del_t \star \del_x = \del_t \del_x$.
The pair of operators $(L,T)$ and the equation (\ref{lax})
are called the {\it Lax pair} and the {\it NC Lax equation},
respectively. 

The NC Lax equation (\ref{lax}) is derived 
from the compatible condition of the following 
NC version of the linear system
\begin{eqnarray}
 &&L\star \psi=\lambda \psi,\\
 &&\fr{\del \psi}{\del t}+T\star \psi=0,
\label{evolution}
\end{eqnarray}
where the eigenvalue $\lambda$ is a constant.
On commutative spaces,
Eq. (\ref{evolution}) is an evolution equation
and the existence of the Lax representation (\ref{lax})
suggests the compatibility of the linear systems.
On NC spaces, however, the RHS of
the equations (\ref{evolution}) could contain
infinite number of derivatives of some variables
and the geometrical meaning might be vague.
Therefore at this stage, the integrability
of the Lax equation is not trivial.
In the next section, we will see a NC
Burgers equation is actually linearizable,
which suggests the NC deformation
would have good properties.
Furthermore, we would like to comment that
the NC (anti-)self-dual Yang-Mills equation
is integrable and derived from the compatibility 
of linear systems with spectral parameters. (e.g. \cite{MaWo})

Now let us construct the NC Burgers equation
with the Lax representation
by the {\it Lax-pair generating technique} \cite{ToYu}.
The technique is a method to
find a corresponding $T$-operator for a given $L$-operator
and based on the following ansatz for the $T$-operator
\begin{eqnarray}
\label{ansatz}
T=\del_i^n L^m +T^\prime.
\end{eqnarray}
Then the problem reduces to that for the $T^\prime$-operator
and becomes enough easy to solve in many cases
including NC cases \cite{Toda, HaTo}.

Let us apply this technique to 
the NC extension of the Burgers equation.
The $L$-operator of the Burgers equation is given by
\begin{eqnarray}
L_{\scr\mbox{Burgers}}=\del_x+u(t,x).
\end{eqnarray}
The ansatz for the $T$-operator is now taken as
\begin{eqnarray}
T_{\scr\mbox{Burgers}}
=\del_x L_{\scr\mbox{Burgers}}+T_{\scr\mbox{Burgers}}^\prime,
\end{eqnarray}
which is the case for $n=1$ in the general ansatz (\ref{ansatz}).
The ansatz for $n=2,3,\ldots$ gives rise to
the Burgers hierarchy, which is discussed in section 4.

Then the NC Lax equation becomes
\begin{eqnarray}
[\del_x+u,T_{\scr\mbox{Burgers}}^\prime]_\star= u_x \del_x+u_t+u_x \star u,
\label{tpr}
\end{eqnarray}
where $u_x:=\del u/\del x$ and so on.
Here the term $u_x \del_x$ in the RHS of Eq. (\ref{tpr})
is troublesome because the Lax equation should be a
differential equation without bare derivatives $\del_i$.
Hence we have to delete this term to find an appropriate
$T^\prime$-operator so that the bare derivative term in
the LHS of Eq. (\ref{tpr})
should be canceled out.
In order to do this, we can take the $T^\prime$-operator as
the following form:
\begin{eqnarray}
T_{\scr\mbox{Burgers}}^\prime=A\del_x+B,
\end{eqnarray}
where $A$ and $B$ are polynomials of $u, u_x, u_t,$ etc.
Then the Lax equation becomes $f\del_x+g=0$ and
the condition $f=0$ determines some part of $A,B$ and
finally a differential equation $g=0$ is expected to be
the Burgers equation which possesses the Lax representation.

The condition $f=0$ is
\begin{eqnarray}
A_x+[u,A]_\star=u_x.    
\end{eqnarray}
The solution is $A=u$.\footnote{Exactly speaking,
the general solution is $A=u+\alpha$ where $\alpha$ is a constant.
However this constant can be absorbed by the scale transformation
$u\rar u+\alpha/(2-2\beta)$.
In this article, we omit such constants.}
Then the differential equation $g=0$ becomes
\begin{eqnarray}
B_x+[u,B]_\star=u_t+u\star u_x+u_x\star u.
\end{eqnarray}
Taking the dimensions of the variables into account,
that is, $[x]=-1, [u]=1, [t]=-2$, hence, $[B]=2$,
we can take the unknown $B$ as\footnote{Here we don't consider
fractional terms such as $u_{xx}\star u^{-1}$ and so on.
This constraint corresponds to $B_2:=(L^2)_{\star\geq 0}$ in the
framework of Sato theory. (cf. section 9.)}
\begin{eqnarray}
B=a u_x+b u^2,
\end{eqnarray}
where $a$ and $b$ are constants.\footnote{
We note that $u\star u\equiv u^2$.}

Finally we get the NC version of the Burgers
equation with parameters:
\begin{eqnarray}
\label{burgers}
u_t-a u_{xx}+(1+a-b)u_x\star u+(1-a-b)u\star u_x=0,
\end{eqnarray}
whose Lax pair is
\begin{eqnarray}
 \left\{
 \ba{lll}
 L_{\scr\mbox{Burgers}}&=&\del_x+u,\\
 T_{\scr\mbox{Burgers}}&=&\del_x^2+2u\del_x+(a+1)u_x+bu^2.
 \ea
 \right.
\end{eqnarray}
In the commutative limit, it reduces to
\begin{eqnarray}
u_t-a u_{xx}+2(1-b)u u_x=0.
\end{eqnarray}

We note that the nonlinear term $uu_x$ in the Burgers equation
should be extended not as symmetric forms 
but as Eq. (\ref{burgers})
so that the NC Burgers equation should possess
the Lax representation.

This parameter family is a general form with Lax representation.
Of course, some parameters can be absorbed 
by a scale transformation. 

However, on NC spaces,
it is not clear whether
the Lax representation has good properties or not
in the integrable sense.
In the next section, 
let us seek for the condition on the constants 
that the Burgers equations should be linearizable.

\section{Noncommutative Cole-Hopf Transformation}

In commutative case, it is well known that
the Burgers equation is linearized by the
Cole-Hopf transformation
\begin{eqnarray}
\label{com_ch}
u=\fr{1}{c}\del_x \log\psi=\fr{1}{c}\fr{\psi_x}{\psi}.
\end{eqnarray}
Taking the transformation for the Burgers equation (\ref{burgers}),
we get\footnote{Here we can set the constant $C(t)$ zero  
in Eq. (\ref{com_linear}) after the linearization 
without loss of generality
because it can be absorbed 
by the scale transformation $\psi \rar \psi \exp
\left\{\pm\int^t C(t^\prime)dt^\prime\right\}$.}
\begin{eqnarray}
\label{com_linear}
\psi_t=a\psi_{xx}
-\left(a-\fr{b-1}{c}\right)\fr{\psi_x^2}{\psi}.
\end{eqnarray}
Hence we can see that only when $ac=b-1$, 
the Burgers equation
reduces to the linear equation $\psi_t=a\psi_{xx}$.\footnote{
Without loss of generality, we can take $a>0$.
Then the linear equation is just the diffusion equation 
or the heat equation
where $a$ shows the coefficient of viscosity.}
The linearizable Burgers equation becomes
\begin{eqnarray}
\label{com_lin_burgers}
u_t-a u_{xx}-2ac u u_x=0.
\end{eqnarray}
We note that the scale transformations $t\rar (1/a)t$
and  $u\rar (1/c)u$ absorb the constants $a$ and $c$
in Eqs. (\ref{com_ch}) and (\ref{com_lin_burgers}), respectively.

This transformation (\ref{com_ch}) 
still works well in NC case.
Then we have to treat the inverse of $\psi$ carefully.
There are typically two possibilities to
define the NC version of the Cole-Hopf transformation:\footnote{
There would be other candidates, such as $u=\psi^{-\alpha}\star
\psi_x\star \psi^{-\beta}$ where $\alpha+\beta=1$. However 
they do not seem to lead to linear equations because
$\del_i\psi^{-\alpha}=-\psi^{-\alpha}\star\del_i\psi^\alpha\star 
\psi^{-\alpha}$ makes it complicated.}
\begin{eqnarray}
\mbox{(i)}&& u=\dis\fr{1}{c}\psi_x\star \psi^{-1}
\label{ch1}\\
\mbox{(ii)}&& u=\dis\fr{1}{c}\psi^{-1}\star
\psi_x\label{ch2}
\end{eqnarray}

\vspace{3mm}

In the case (i),
we can see that when $a+b=1, c=-1$,\footnote{
Here we omit the possibility: $a=0, b=1$ 
because the NC Burgers equation (\ref{burgers}) 
becomes trivial in this case.}
the NC Burgers equation reduces to 
the equation: $(\del_x-\psi_x\star \psi^{-1})
\star(\psi_t-a\psi_{xx})=0$.
Hence the solutions of the NC diffusion 
equation\footnote{Here we can also put $a>0$ 
as in commutative case.}
\begin{eqnarray}
\label{diffusion}
\psi_t=a\psi_{xx},
\end{eqnarray}
give rise to the exact solutions of the NC Burgers equation via
the NC Cole-Hopf transformation (\ref{ch1}).
The naive solution of 
the NC diffusion equation (\ref{diffusion}) is
\begin{eqnarray}
\label{naive_sol}
\psi(t,x)=1+\sum_{i=1}^N h_i e^{a k_i^2t}\star e^{\pm k_i x}
=1+\sum_{i=1}^N h_i e^{\fr{i}{2}a k_i^3\theta}e^{a k_i^2t\pm k_i x},
\end{eqnarray}
where $h_i, k_i$ are complex constants.
The final form in (\ref{naive_sol})
shows that the naive solution on commutative space
is deformed by $e^{\fr{i}{2}a k_i^3\theta}$ 
due to the noncommutativity.
This reduces to the $N$-shock wave solution in fluid
dynamics. Hence we call it the {\it NC N-shock wave solution}.
The explicit representation in terms of $u$
is hard to obtain because the derivation of $\tau^{-1}$
is non-trivial.
However we can discuss the asymptotic behaviors
at $t\rar \pm \infty$. The effect of the NC deformation
is easily seen.
In fact , exact solutions for $N=1,2$ 
are obtained by L.~Martina and O.~Pashaev 
\cite{MaPa} and nontrivial effects of the NC-deformation
are actually reported.

If we want to know more general solutions,
it would be appropriate to take the Fourier transformation
under some boundary conditions.
The calculation is the same as the commutative case.
The initial value problem is also well-defined,
that is, the initial condition
$u(t=0,x)=-\psi_x\star \psi^{-1}\vvert_{t=0}$
is an appropriate one.

Let us comment on multi-soliton solutions with no scattering process.
Defining $z:=x+vt, \zb:=x-vt$, we easily see
\begin{eqnarray}
 f(z)\star g(z)= f(z) g(z)
\end{eqnarray}
because the star-product (\ref{star_11}) is rewritten
in terms of $(z,\zb)$ as
\begin{eqnarray}
 f(z,\zb)\star g(z,\zb)=
e^{iv\theta\left(
\partial_{\zb^\prime}
\partial_{z^{\prime\prime}}-
\partial_{z^\prime}
\partial_{\zb^{\prime\prime}}
\right)}f(z^\prime,\zb^\prime) g(z^{\prime\prime},\zb^{\prime\prime})
\Big{\vert}_{\scr
\ba{c}
z^{\prime}
=z^{\prime\prime}=z\\
\zb^{\prime}
=\zb^{\prime\prime}=\zb.
\ea}
\end{eqnarray}
This situation is realized when all $k_i$ 
are the same: $k_1=k_2=\cdots=k_N=k(=:v/a)$
including one-soliton solutions.
The NC one shock-wave solution \cite{MaPa} is essentially
the same as the commutative one because of the above observation.
In fact, their one shock-wave solution is reduced to
our solution (\ref{naive_sol}) by putting $k_1=0$ in \cite{MaPa}.
The condition $k_1=0$ is taken without loss of generality
and then the effect of NC-deformation disappears.

We note that the solution $\psi^{\scr{\mbox{sol}}}$ of 
the diffusion equation (\ref{diffusion}) do not yield
all solutions of the NC Burgers equation
because the NC Cole-Hopf transformation (\ref{ch1})
is a one-way map.
However the transformation $\psi^{\scr{\mbox{sol}}}\rar g_\psi\star 
\psi^{\scr{\mbox{sol}}}$ gives rise to the solution of
the directly transformed equation $(\del_x-\psi_x\star \psi^{-1})
\star(\psi_t-a\psi_{xx})=0$ from the NC Burgers equation,
where $g_\psi$ is the so-called {\it NC transition operator}
which satisfies $\del_x g_\psi= (\psi_x\star \psi^{-1})\star g_\psi $.
The existence of $g_\psi$ would be guaranteed \cite{NC}
and in principle we can construct all solutions of
the NC Burgers equation via the inverse of the 
NC Cole-Hopf transformation.


\vspace{3mm}

In the case (ii),
the same discussion leads us to the similar conclusion
that when $a-b=-1, c=1$,\footnote{
Here we also omit the possibility: $a=0, b=1$ 
for the same reason as in the case (i).}
the solutions of the NC diffusion equation (\ref{diffusion})
yields the exact solutions of the NC Burgers equation via
the NC Cole-Hopf transformation (\ref{ch2}).

\begin{figure}[htbn]
\begin{center}
\includegraphics[width=60mm]{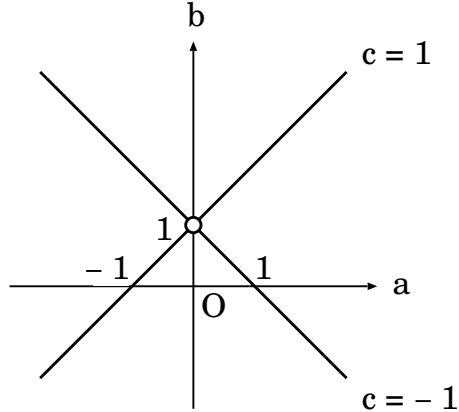}
\caption{The region where the NC Burgers equation
can be linearized}
\label{region}
\end{center}
\end{figure}

\vspace{3mm}
\noindent
The region where the NC Burgers equation (\ref{burgers})
can be linearized is shown in Fig. \ref{region}.

It is interesting that the condition on $a,b$ is equivalent to
that each part of two coefficients of $u_x\star u$ and $u \star u_x$
in the NC Burgers equation (\ref{burgers}) vanishes.
The result is summarized in Table 1.

\begin{center}
Table 1: The Linearizable NC Burgers Equation\\
\begin{tabular}{|c|c|c|} \hline
~&NC Cole-Hopf transformation&NC Burgers Equation\\\hline\hline
(i)&$u=-\psi_x\star \psi^{-1}$
&$u_t-a u_{xx}+2a u_x\star u=0$\\\hline
(ii)&$u=\psi^{-1}\star\psi_x$
&$u_t-a u_{xx}-2a u\star u_x=0$\\\hline
\end{tabular}\\
\end{center}

This is formally consistent with the condition that
the matrix Burgers equation should be integrable \cite{OlSo},
which would be reasonable because the variable $u$
in the NC deformed Burgers equation can be rewritten
as the infinite-size matrix by the Weyl transformation.
Of course, the notions of time evolution are different.

In the commutative limit, the linearizable NC Burgers equation
reduces to commutative one (\ref{com_lin_burgers})
with $c=\pm 1$.

\section{Conserved Quantities of the Noncommutative Burgers Equation}

Here we would like to comment on conserved quantities
of NC Burgers equation.
The discussion is basically the same as commutative case
because both the differentiation and the integration
are the same as commutative ones in the Moyal representation.

Let us suppose the conservation law
\begin{eqnarray}
{\partial J(t,x)\over{\partial t}}={\partial K(t,x)\over{\partial x}}.
\end{eqnarray}
then the conserved quantity is given by an integral
\begin{eqnarray}
Q(t)=\int_{-\infty}^{\infty}J(t,x)dx.
\end{eqnarray}
The proof is straightforward:
\begin{eqnarray}
{d Q\over{dt}}={\partial
\over{\partial t}}\int_{-\infty}^{\infty}J(t,x)dx
=\int_{-\infty}^{\infty}{\partial J(t,x)\over{\partial
t}}dx=\int_{-\infty}^{\infty}{\partial K(t,x)\over{\partial x}}dx =0,
\end{eqnarray}
unless the integrand $K(t,x)$ vanishes or is periodic at spatial infinity.
The convergence of the integral is also expected because
the star-product naively reduces to the ordinary product 
at spatial infinity due to: $\del_x \sim {\cal{O}}(x^{-1})$.

\vspace{3mm}

On commutative spaces,
the existence of infinite number of conserved quantities would 
lead to infinite-dimensional hidden symmetry from Noether's theorem.
In Liouville sense, the existence is necessary condition for
complete integrability unlike dynamical systems
with finite-dimensional degree of freedom.

On NC spaces, this is also true and
the existence of infinite number of conserved quantities 
would be meaningful.
Many NC field equations with infinite number of conserved quantities
have been found by the bi-complex method \cite{DiMH, DiMH2, GrPe}.
The bi-complex method also seems to be applicable to 
the NC Burgers equation.
However this time, it is not so trivial whether 
Noether's theorem is valid or not.


\section{Reduction from the Noncommutative
(Anti-)Self-Dual Yang-Mills Equation}

On commutative spaces,
there is a famous conjecture, {\it Ward conjecture} \cite{Ward}.
The statement is that almost all lower-dimensional 
integrable equations can be derived from 
(anti-)self-dual Yang-Mills equation by reductions.
This conjecture is almost confirmed now \cite{MaWo}.

It is very interesting to study whether
this conjecture still holds on NC spaces or not.
In this section, we show that the NC Burgers
equation could be derived from a NC 
(anti-)self-dual Yang-Mills equation by reduction,
which is one example of NC Ward conjecture \cite{HaTo}.

Let us consider the following NC
(anti-)self-dual Yang-Mills equation with $G=U(1)$
(Eq. (3.1.2) in \cite{MaWo}):
\begin{eqnarray}
 &&\del_{w} A_z -\del_{z} A_w+[A_w,A_z]_\star =0,~~~
 \del_{\tilde{w}} A_{\tilde{z}} -\del_{\tilde{z}} A_{\tilde{w}}
 +[A_{\tilde{w}},A_{\tilde{z}}]_\star =0,\nn
 &&\del_{z} A_{\tilde{z}} -\del_{\tilde{z}} A_{z}
 +\del_{\tilde{w}} A_{w} -\del_{w} A_{\tilde{w}}
 +[A_{\tilde{z}},A_z]_\star
 +[A_{w},A_{\tilde{w}}]_\star=0.
\label{asdym}
\end{eqnarray}
where $(z,\tilde{z},w,\tilde{w})$ and $A_{z,\tilde{z},w,\tilde{w}}$ 
denote the coordinates 
of the original $(2+2)$-dimensional space and
the gauge fields, respectively. 
We note that the commutator part should remain
though the gauge group is $U(1)$ because the elements of the
gauge group could be operators and the gauge group could be
considered to be non-abelian: $U(\infty)$.
This commutator part actually plays an important role
as usual in NC theories. 

Now let us take the simple dimensional reduction $\del_{\tilde{z}}
=\del_{\tilde{w}}=0$ and put the following constraints:
\begin{eqnarray}
 A_{\tilde{z}}=A_{\tilde{w}}=0,~~~A_z=u,~~~A_w=a u_z+(1-b)u^2.
\end{eqnarray}
Then the NC (anti-)self-dual Yang-Mills 
equation (\ref{asdym}) reduces to
\begin{eqnarray}
 u_w-a u_{zz}+(1+a-b) u_z\star u +(1-a-b) u\star u_z=0. 
\end{eqnarray}
This is just the NC Burgers equation (\ref{burgers})
with $w\equiv t,~z\equiv x$.
We note that without the commutator part $[A_w,A_z]_\star$,
the nonlinear term should be symmetric: $(u_z\star u + u\star u_z)$,
which cannot lead to the Lax representation as is commented below
Eq. (\ref{burgers}). 
This shows that the special feature in NC gauge theories
plays a crucial role. Therefore the NC Burgers
equation is expected to have some non-trivial property special to
NC spaces such as the existence of $U(1)$ instantons.

Essentially the same argument is presented in \cite{MaPa}
from a different viewpoint.

\section{Noncommutative Hierarchy Equations}

Now let us look for the Lax representations
of the NC Burgers equation with the higher-dimensional
time evolution by the Lax-pair generating technique:
\begin{eqnarray}
 \left[L_{\scr\mbox{Burgers}},
 T_{n\scr\mbox{th-h}}+\del_{t_n}\right]_\star=0,
\label{hie}
\end{eqnarray}
where the dimensions are given 
by $[t_n]=-n, [T_{n\scr\mbox{th-h}}]=n$
and the noncommutativity could be introduced as $[t_n,x]=i\theta_n$.
The Lax representations (\ref{hie})
is derived from the compatible conditions of the
NC linear systems:
\begin{eqnarray}
 &&L_{\scr\mbox{Burgers}}\star \psi=\lambda \psi,\\
 &&\fr{\del\psi}{\del t_n}+T_{n\scr\mbox{th-h}}\star \psi=0.
\label{evolution_hie}
\end{eqnarray}
This time, Eq. (\ref{evolution_hie}) is not an evolution
equation. However as the previous discussion,
some geometrical meaning would be expected.
Then, the existence of infinite number of hierarchy equations
would suggest infinite-dimensional hidden symmetry
which is expected to be deformed symmetry from commutative one.

Now let us take the other ansatz 
for the operator $L_{\scr\mbox{Burgers}}=\del_x+u$ as
\begin{eqnarray}
T_{(n+1)\scr\mbox{th-h}}=\del_x^n L_{\scr\mbox{Burgers}} 
+T_{(n+1)\scr\mbox{th-h}}^\prime.
\end{eqnarray}
Then the unknown part is reduced 
to $T_{(n+1)\scr\mbox{th-h}}^\prime$
which is determined so that Eq. (\ref{hie})
is a differential equation.
The results are as follows.

\begin{itemize}

\item For $n=1$, the NC Lax equation gives
the (second-order) NC Burgers equation
(\ref{burgers}).

\item For $n=2$, the NC Lax equation gives
the third-order NC Burgers equation.

The Lax pair is given by
\begin{eqnarray}
L_{\scr\mbox{Burgers}} =\del_x+u(t,x),~~~
T_{\scr\mbox{3rd-h}}=\del_x^2 L_{\scr\mbox{Burgers}}
+T_{\scr\mbox{3rd-h}}^\prime,
\end{eqnarray}
where
\begin{eqnarray}
\del_x^2 L_{\scr\mbox{Burgers}}=\del_x^3+u\del_x^2 +2u_x \del_x+u_{xx}.
\end{eqnarray}
Substituting this ansatz into the NC Lax equation,
we can take more explicit form on
$T_{\scr\mbox{3rd-h}}^\prime$ as
\begin{eqnarray}
T_{\scr\mbox{3rd-h}}^\prime=A\del_x^2+B\del_x+C,
\end{eqnarray}
where $A,B$ and $C$ are polynomials of $u, u_x, u_t,$ etc.
In the similar way to the $n=1$ case,
the unknown polynomials satisfy the following differential
equations
\begin{eqnarray}
&&A_x+[u,A]_\star-2u_x=0,\nn
&&B_x+[u,B]_\star-2A\star u_x-u_{xx}-2u_x\star u=0,\nn
&&C_x+[u,C]_\star-A\star u_{xx}-B\star u_x-u_t
-2u_x^2-u_{xx}\star u =0,
\label{third}
\end{eqnarray}
and the solutions are:
\begin{eqnarray}
&&A=2u,~~~B=u_x+3u^2,\nn
&&C=au_{xx}+bu_x\star u +cu\star u_x+du^3,
\end{eqnarray}
where the coefficients $a,b,c$ and $d$ are constants.

Then the last equation of (\ref{third})
yields the third-order NC Burgers equation with parameters:
\begin{eqnarray}
\label{3_hie}
&&u_t-au_{xxx}+(1+a-b)u_{xx}\star u+(2-a-c)u\star u_{xx}
+(3-b-c)u_x^2\nn
&&+(b-d)u_x\star u^2+(c-b-d)u\star u_x \star u 
+(3-c-d)u^2\star u_x=0,
\end{eqnarray}
whose Lax pair is
\begin{eqnarray}
 \left\{
 \ba{lll}
 L_{\scr\mbox{Burgers}}&=&\del_x+u,\\
 T_{\scr\mbox{3rd-h}}&=&\del_x^3+3u\del_x^2+3(u_x+u^2)\del_x
  +(a+1)u_{xx}+bu_x\star u +cu\star u_x+du^3.
 \ea
 \right.
\end{eqnarray}
The parameter family of this equation 
formally coincides with four integrable equations
given in Theorem 3.6 in \cite{OlSo}, that is,
two type of the third-order NC Burgers equations and 
two type of NC modified KdV equations
up to scale transformations:

\begin{itemize}

\item the third-order NC Burgers equation: ($a=-1,b=c=d=0$)
\begin{eqnarray}
 u_t+u_{xxx}+3u\star u_{xx}+3u_x^2+3u^2\star u_x=0.
\label{2_burgers}
\end{eqnarray}

\item the third-order (conjugated) NC Burgers equation: ($a=-1,b=c=3,d=0$)
\begin{eqnarray}
 u_t+u_{xxx}-3u_{xx}\star u-3u_x^2+3u_x\star u^2=0.
\end{eqnarray}

\item NC modified KdV equation via NC Miura map from NC KdV equation
\cite{DiMH2}: ($a=-1, b=0, c=d=3$)
\begin{eqnarray}
\label{mKdV1}
 u_t+u_{xxx}-3u_x\star u^2-3u^2\star u_x=0.
\end{eqnarray}
Our result gives the Lax representation of the Miura-mapped
NC KdV equation.

\item NC modified KdV equation: ($a=-1, c=0, b=d=3$)
\begin{eqnarray}
\label{mKdV2}
 u_t+u_{xxx}+3[u,u_{xx}]_\star-6u\star u_x\star u=0.
\end{eqnarray}
This has another Lax representation:
\begin{eqnarray}
  \left\{
 \ba{lll}
 L=\del_x^2+2u\del_x,\\
 T=4\del_x^3+12u\del_x^2+6(u^2+u_x)\del_x.
 \ea
 \right.
\end{eqnarray}

\end{itemize}

Let us comment on the linearizability.
This time, the linearizable condition by the NC 
version of Cole-Hopf transformation 
leads to the restricted situation $a=0,b=1,c=2,d=1$
where the third-order NC Burgers equation
(\ref{3_hie}) becomes trivial.
The result shows that the linearizable condition
is too strict for the third-order NC Burgers equation
(\ref{3_hie}) due to the nonlinear effect.

\item For $n=3$, the NC Lax equation gives
the fourth-order NC Burgers equation. 

The Lax pair is given by
\begin{eqnarray}
L_{\scr\mbox{Burgers}} =\del_x+u(t,x),~~~
T_{\scr\mbox{4th-h}}=\del_x^3 L_{\scr\mbox{Burgers}}
+T_{\scr\mbox{4th-h}}^\prime.
\end{eqnarray}
Substituting this ansatz into the Lax equation (\ref{lax}),
we can take more explicit form on
$T_{\scr\mbox{4th-h}}^\prime$ as
\begin{eqnarray}
T_{\scr\mbox{4th-h}}^\prime=A\del_x^3+B\del_x^2+C\del_x+D,
\end{eqnarray}
where $A,B,C$ and $D$ are polynomials of $u, u_x, u_t,$
etc
and are determined 
in the similar way to the cases for $n=1,2$ as
differential equations
\begin{eqnarray}
A&=&3u,~~~B=3u_x+6u^2,\nn
C&=&u_{xx}+4u_x\star u +8u\star u_x+4u^3,\nn
D&=&au_{xxx}+bu_{xx}\star u+cu\star u_{xx}+du_x^2\nn
&&+eu_x\star u^2+fu\star u_x\star u+gu^2\star u_x+hu^4,
\end{eqnarray}
where the coefficients $a,b,\ldots,h$ are constants.
Then we can get the fourth-order NC Burgers
equation with parameters:
\begin{eqnarray}
&&u_t-au_{xxxx}+(1+a-b)u_{xxx}\star u+(3-a-c)u\star
u_{xxx}\nn
&&+(4-b-d)u_{xx}\star u_x+(6-c-d)u_{x} \star u_{xx}\nn
&&+(b-e)u_{xx}\star u^2+(c-b-f)u \star u_x \star
u+(6-c-g)u^2\star 
u_{xx}\nn
&&+(d-e-f)u_x^2\star u+(4-e-g)u_x\star u\star u_x
+(8-d-f-g)u\star u_x^2\nn
&&+(e-h)u_x\star u^3+(f-e-h)u\star u_x\star u^2\nn
&&+(g-f-h)u^2\star u_x\star u+(4-g-h)u^3\star u_x=0.
\end{eqnarray}
\end{itemize}

In this way, we can generate 
the higher-order NC Burgers equations.
The ansatz for the $(n+1)$-th order is more explicitly given by
\begin{eqnarray}
T_{(n+1)\scr\mbox{-th}}=\del_x^n L +T_{(n+1)\scr\mbox{-th}}^\prime
=\del_x^{n+1}+\sum_{k=0}^{n}
\fr{n!}{k!(n-k)!}(\del^k_x u )\del_x^{n-k}
+\sum_{l=0}^{n} A_l \del_x^{n-l},
\end{eqnarray}
where $A_l$ are homogeneous polynomials of $u,u_x,u_{xx}$
and so on, whose degrees are $[A_l]=l+1$. 
These unknown polynomials are determined 
one by one as $A_0=nu$ and so on.

\section{Sato's Approach to the Noncommutative Hierarchy}

In this final section,
we present NC versions of
the Burgers equation and the Burgers hierarchy
in the framework of the Sato theory \cite{SaSa}
by using the pseudo-differential operator.
We look for the Lax representation of the NC Burgers hierarchy.

Let us introduce the following Lax operator
as a pseudo-differential operator:
\begin{eqnarray}
 L = \partial_x + u_1 + u_2 \partial_x^{-1} 
 + u_3 \partial_x^{-2} + u_4 \partial_x^{-3} + \cdots,
\end{eqnarray}
where the infinite number of fields $u_m ~(m=1,2,~\ldots)$
depend on infinite number of variables
$(t_1,t_2,t_3,~\ldots)$.
The action of the operator $\del_x^n$ on
a multiplicity function $f(x)$ is given by
\begin{eqnarray}
 \del_x^{n}\cdot f:=\sum_{i\geq 0}
\left(\ba{c}n\\i\ea\right)
(\del_x^i f)\del^{n-i},
\end{eqnarray}
where the binomial coefficient is defined as
\begin{eqnarray}
 \left(\ba{c}n\\i\ea\right):=\fr{n(n-1)\cdots (n-i+1)}{i(i-1)\cdots 1}.
\end{eqnarray}
We note that the definition can be extended to the negative
powers of $\del_x$.
The examples are,
\begin{eqnarray*}
 \del_x^{-1}\cdot f&=& f\del_x^{-1}-f_x\del_x^{-2}+f_{xx}\del_x^{-3}-\cdots,\\
 \del_x^{-2}\cdot f&=& 
f\del_x^{-2}-2f_x\del_x^{-3}+3f_{xx}\del_x^{-4}-\cdots,\\
 \del_x^{-3}\cdot f&=& 
f\del_x^{-3}-3f_x\del_x^{-4}+6f_{xx}\del_x^{-5}-\cdots,
\end{eqnarray*}
where $\del_x^{-1}$ in the RHS
acts as an integration operator $\int^x dx$.
For more on foundation of the pseudo-differential operators 
and the Sato theory, see e.g. \cite{DJM, Blaszak, Kupershmidt}.

The noncommutativity for them can be introduced arbitrarily.
Thus we do not fix the noncommutativity here.
At the end of the present section, we comment on this point. 

The Lax representation 
for the NC Burgers hierarchy in Sato's framework is given by
\begin{eqnarray}
 \left[\del_{t_m}-B_m, L\right]_\star=0,
\label{lax_sato}
\end{eqnarray}
where $B_m$ is given here by
\begin{eqnarray}
 B_m:=(\unb{L\star \cdots \star L}_{m{\scr\mbox{ times}}})_{\geq 1}
=:(L^m)_{\star\geq 1}.
\end{eqnarray}
The suffix ``$\geq 1$'' represents the positive power part of $L^m$.
A few concrete examples are as follows:
\begin{eqnarray}
 B_1&=& \del_x,\nonumber\\
 B_2  &=& \partial_x^2 + 2u_1\partial_x,\nonumber\\ 
 B_3  &=& \del_x^3 +3 u_1 \del_x^2 + 3(u_2 +  u_1^2 +  (u_1)_x)\del_x.
\end{eqnarray}

The replacement of the products of fields in the commutator
with the star products means the NC extension.
In this approach, the geometrical meaning of the Lax representation
is also vague. However we can expect that the Lax equations
actually contain integrable equations as in the previous sections.

Now let us discuss the existence of
the NC Burgers hierarchy.
The NC Burgers hierarchy could be 
derived by putting the following constraint for
the Lax equations (\ref{lax_sato})
\begin{eqnarray}
 L=B_1~(=:L_{\scr\mbox{Burgers}}),
\end{eqnarray}
which implies
\begin{eqnarray}
 u_k=0,~~~k=2,3,4,~\ldots.
\end{eqnarray}
This means that the Lax equations (\ref{lax_sato})
can be represented in terms of one kind of field $u_1\equiv u$,
which guarantees the existence of the hierarchy.
Now we can see that the Lax equation (\ref{lax_sato})
just gives a differential equation.

The hierarchy equations are as follows:
\begin{itemize}
\item For $m=1$, the Lax equation (\ref{lax_sato})
reduces to $u_{t_1}=u_x$, which means $t_1=x$.

\item For $m=2$, the Lax equation (\ref{lax_sato}) becomes 
the second order NC Burgers equation
\begin{eqnarray}
\label{second}
 \fr{\del u}{\del t_2} 
= [B_2,L_{\scr\mbox{Burgers}}]_\star
= [\del_x^2+2u\del_x, \del_x +u]_\star
 =u_{xx}+2u\star u_x.
\end{eqnarray}
This is just one of the linearizable 
NC Burgers equation with $t_2\equiv t$. (See Table 1.)

\item For $m=3$, the Lax equation (\ref{lax_sato}) is
the third order NC Burgers equation
\begin{eqnarray}
 \fr{\del u}{\del t_3}&=&[B_3,L_{\scr\mbox{Burgers}}]_\star\nn
 &=&
 [\del_x^3+3\del_x^2+3(u^2+u_x)\del_x, \del_x +u]_\star\nn
 &=&u_{xxx}+3u\star u_{xx}+3u_x^2 +3u^2\star u_{x}.
\end{eqnarray}
This just coincides with 
the third order NC Burgers equation (\ref{2_burgers}).

\item For $m=4$, the Lax equation (\ref{lax_sato}) is
the fourth order NC Burgers equation
\begin{eqnarray}
 \fr{\del u}{\del t_4}&=&
 [\del_x^4+4u\del_x^3+6(u^2+u_x)\del_x^2
 +4(u^3+u_x \star u+2u \star u_x)\del_x,
 \del_x+u]_\star\nn
 &=&u_{xxxx}+4u\star u_{xxx}+4u_{xx}\star u_x
 +6u_x \star u_{xx}\nn
 &&+6u^2\star u_{xx}+4u_x u u_x +8u\star u_x^2 +4u^3\star u_x.
\end{eqnarray}
\end{itemize}

In this way, we can obtain infinite number of
NC hierarchy equations which possess
the Lax representations with no parameter.
The results for $m=2,3$ suggest that this approach
gives rise to integrable equations directly.
The relationship between this approach and 
the Lax-pair generating technique 
is $B_m=(T_{m\scr\mbox{th-h}})_{\geq 1}$.
Hence the Lax-pair generating technique
would yield wider class of integrable equations
than Sato's approach such as the modified KdV equations 
(\ref{mKdV1}) and (\ref{mKdV2}).

\vspace{3mm}

Let us comment how to introduce the noncommutativity of
infinite-dimensional parameter spaces: $[t_i,t_j]=i\theta^{ij}$.
A natural one is $\theta^{12}\neq 0,\theta^{34}\neq 0,\ldots,
\theta^{2n-1,2n}\neq 0,\ldots$ and otherwise $=0$.
However the noncommutativity $\theta^{2n-1,2n}\neq 0$ 
introduces infinite number of derivatives with respect to
$t_{2n-1}, t_{2n}$.
If we respect the notion of time evolutions, it is reasonable to 
take the special possibility: $\theta^{12}\neq 0$ and otherwise $=0$.
Then problematic directions are $t_1$ and $t_2$-directions only.
However, we know that the NC Burgers equation is actually
integrable and 
the integrability is guaranteed in $t_1,t_2$-direction.
In the other directions, time evolution is well-defined
and the hierarchy equations can be considered as
infinite-side matrix evolution equations 
because of $\hat{u}=\sum_{m,n=0}^\infty u_{mn}(t_3,t_4,\ldots)
\vert m\ket\bra n\vert$.
Then the situation belongs to infinite-dimensional version of 
matrix Sato theory and various discussions would be possible.
This time, the results in \cite{OlSo} are applicable to
the present discussion and the third order NC Burgers equation
is proved to be integrable.

\section{Conclusion and Discussion}

In this article, we gave NC versions
of the Burgers equation and the Burgers hierarchy
which possess the Lax representations.
We proved that the NC Burgers equation
is linearizable by the NC versions of the Cole-Hopf 
transformations (\ref{ch1}) and (\ref{ch2}).
The linearized equation is the (NC) diffusion equation
of first order with respect to time
and the initial value problem is well-defined.
We obtained the exact solutions of the linearized equation
and discussed the effects of the NC deformation.
Furthermore, we rederived the NC Burgers equation
from both NC (anti-)self-dual Yang-Mills equation,
which would be an evidence of 
NC Ward conjecture.
Finally we showed the NC Burgers hierarchy
in the frameworks of the Lax-pair generating technique
and the Sato theory,
which would lead to
the completion of NC Sato theory.

The results show that the NC Burgers equation 
is completely integrable even though the
NC Burgers equation contains infinite number
of time derivatives in the nonlinear term.
The linearized equation is the (NC) diffusion equation
of first order with respect to time
and the initial value problem is well-defined.
This is a surprising discovery and 
a good news for the study of 
NC extension of integrable equations.
There would be many further directions
such as NC extension of the Hirota's bilinear method
which is a simple generalization of the Cole-Hopf 
transformation.
The existence of NC hierarchies 
for other integrable equations
have been already proved in \cite{Toda, appear}.
Hence if we succeed in the bilinearization,
the NC Sato theory must be constructed.
Then we can discuss the structure of the solution spaces
and the symmetry underlying 
the integrability.
The discretization of the
NC Burgers equation is also
considerably interesting in recent developments
of integrable systems.\footnote{The semi-discretization
of the Burgers equation is discussed in detail,
for example, by T.~Tsuchida \cite{Tsuchida}.}
The further study is worth investigating.

\vskip7mm\noindent
{\bf Acknowledgments}
\vskip2mm

\noindent
We would like to thank the YITP at Kyoto University
during the YITP workshops YITP-W-02-04 on ``QFT2002''
and
YITP-W-02-16 on ``Development of Superstring Theory''
where this work has been developed.
We are also grateful to M.~Kato, I.~Kishimoto, A.~Nakamula,
T.~Tsuchida, M.~Wadati and anonymous referees for useful comments.
The work of M.H. was supported in part
by the Japan Securities Scholarship Foundation (\#12-3-0403)
and JSPS Research Fellowships for Young Scientists (\#15-10363).
That of K.T. was financially supported by the Sasagawa Scientific
Research Grant (\#13-089K) from The Japan Science Society
and Grant-in-Aid for Scientific Research (\#15740242).

\begin{appendix} 

\section{Equivalence between the star-product
formalism and the operator formalism}
\vspace{2mm}

The two formalisms of NC field theories,
the star-product formalism and the operator formalism
are equivalent and connected
by the Weyl transformation. 
The Weyl transformation transforms the field $f(x^1,x^2)$ 
in the star-product formalism 
into the infinite-size matrix $\fh(\xh^1,\xh^2)$ as
\begin{eqnarray}
\label{weyl1}
\fh(\xh^1,\xh^2)&:=&\fr{1}{(2\pi)^2}\int dk_1
dk_2~\widetilde{f}(k_1,k_2)
e^{-i(k_1\xh^1+k_2\xh^2)},
\end{eqnarray}
where
\begin{eqnarray}
\label{weyl2}
\widetilde{f}(k_1,k_2)&:=&\int dx^1dx^2~f(x^1,x^2)
e^{i(k_1x^1+k_2x^2)}.
\end{eqnarray}
This map is the composite of twice Fourier transformations
replacing the commutative coordinates $x^1,x^2$ in the exponential
with the NC coordinates $\xh^1,\xh^2$ 
in the inverse transformation:
\begin{eqnarray*}
\begin{array}{ccl}
~&~&f(x^1,x^2)\\
~&\swarrow&~~~~~\, |\\
\widetilde{f}(k_1,k_2)&~&\mbox{Weyl transformation}\\
~&\searrow&~~~~\dar\\
~&~&\fh(\xh^1,\xh^2).
\end{array}
\end{eqnarray*}
The Weyl transformation preserves the product:
\begin{eqnarray}
\widehat{f\star g}=\fh\cdot\hat{g}.
\end{eqnarray}
The inverse transformation of the Weyl transformation
is given directly by
\begin{eqnarray}
f(x^1,x^2)=\int dk_2~e^{-ik_2 x^2}\Bra
x^1+\fr{k_2}{2}\vvert 
\fh(\xh^1,\xh^2)
\vvert x^1-\fr{k_2}{2}\Ket.
\end{eqnarray}
The transformation also maps the derivation and 
the integration one-to-one              . 
Hence the field equation and the solution are also
transformed one-to-one. 
The correspondences are the following (See e.g. \cite{Harvey}):
\vspace{3mm}
\begin{eqnarray*}
\begin{array}{ccc}
\mbox{\fbox{The star-product formalism}}&\lar 
\mbox{Weyl transformation}\rar&
\mbox{\fbox{The operator formalism}}\\\\
\mbox{ordinary
functions}&\mbox{[field]}&\mbox{infinite-size matrices}\\
f(x^1,x^2)&~&\dis\fh(\xh^1,\xh^2)
=\sum_{m,n=0}^{\infty}f_{mn}\vert m\ket\bra n\vert\\\\
\mbox{star-products}&
\mbox{[product]}&
\mbox{multiplications of matrices}\\
\left(f\star(g\star h)=(f\star g)\star h\right)
&\mbox{(associativity)}&\left(\hat{f}(\hat{g}\hat{h})
=(\hat{f}\hat{g})\hat{h}~\mbox{(trivial)}\right)\\\\
{[x^i,x^j]}_\star=i\theta^{ij}&\mbox{[noncommutativity]}
&[\hat{x}^i,\hat{x}^j]=i\theta^{ij}\\\\
\del_if&\mbox{[derivation]}&\del_i\fh:=
[\unb{-i(\theta^{-1})_{ij}\xh^j}_{=:~\delh_i},\fh]\\
\left(\mbox{especially,}~\del_i
x^j=\delta_i^{~j}\right)&~&
\left(\mbox{especially,}~
\del_i\xh^j=\delta_i^{~j}\right)\\\\
\dis\int
dx^1dx^2~f(x^1,x^2)&\mbox{[integration]}&2\pi\theta
{\mbox{Tr}}_{\cH}\fh(\xh^1,\xh^2)\\\\
\ba{c}
\dis\sqrt{\fr{n!}{m!}}
\left(2r^2/\theta\right)^{\fr{m-n}{2}}e^{i(m-n)\varphi}\times\\
2(-1)^nL_n^{m-n}(2r^2/\theta)e^{-\fr{r^2}{\theta}}
\ea
&\mbox{[matrix element]}
&\vert n\ket\bra m\vert
\end{array}
\end{eqnarray*}
\vspace{3mm}

\noindent
where $(r,\varphi)$ is the usual polar coordinate
($r=\left\{(x^1)^2+(x^2)^2\right\}^{\half}$)
and $L^{\alpha}_n(x)$ is the Laguerre polynomial:
\begin{eqnarray}
L_n^\alpha(x):=\fr{x^{-\alpha}e^{x}}{n!}\left(\fr{d}{dx}\right)^n
(e^{-x}x^{n+\alpha}).
\end{eqnarray}

\end{appendix}

\end{document}